\documentclass[prl,aps,preprint,superscriptaddress]{revtex4-1}
\usepackage{graphics}
\usepackage{epsfig}
\usepackage{amsmath}
\usepackage{amsfonts}
\usepackage{amssymb}
\usepackage{graphicx}
\usepackage{color}
\usepackage{tabularx}
\usepackage{txfonts}

\begin{document}
\title{Revisiting Goodenough-Kanamori rules in a new series of double perovskites LaSr$_{1-x}$Ca$_x$NiReO$_6$}
\author{Somnath Jana}
\email[Corresponding author: ] {sj.phys@gmail.com}
\altaffiliation{Present Address: Institute for Methods and Instrumentation in Synchrotron Radiation Research FG-ISRR, Helmholtz-Zentrum Berlin f\"ur Materialien und Energie, Albert-Einstein-Strasse 15, 12489 Berlin, Germany}
\affiliation{Centre for Advanced Materials, Indian Association for the Cultivation of Science, Jadavpur, Kolkata 700032, India }
\affiliation{Department of Physics and Astronomy, Uppsala University, 752 36 Uppsala, Sweden}
\author{Payel Aich}
\affiliation {Department of Materials Science, Indian Association for the Cultivation of Science, Jadavpur, Kolkata 700032, India}
\author{P. Anil Kumar}
\altaffiliation{Present Address: Seagate Technology, 1 Disc Drive, Springtown, Northern Ireland BT48 0BF, United Kingdom}
\affiliation{Department of Engineering Sciences, Uppsala University, 752 36 Uppsala, Sweden}
\author{O.~K.~Forslund}
\affiliation{Department of Applied Physics, KTH Royal Institute of Technology, SE-164 40 Stockholm Kista, Sweden}
\author{E. Nocerino}
\affiliation{Department of Applied Physics, KTH Royal Institute of Technology, SE-164 40 Stockholm Kista, Sweden}
\author{V.~Pomjakushin}
\affiliation{Laboratory for Neutron Scattering \& Imaging, Paul Scherrer Institute, CH-5232 Villigen PSI, Switzerland}
\author{M.~M\aa{}nsson}
\affiliation{Department of Applied Physics, KTH Royal Institute of Technology, SE-164 40 Stockholm Kista, Sweden}
\author{Y.~Sassa}
\affiliation{Department of Physics and Astronomy, Uppsala University, 752 36 Uppsala, Sweden}
\author{Peter Svedlindh}
\affiliation{Department of Engineering Sciences, Uppsala University, 752 36 Uppsala, Sweden}
\author{Olof Karis}
\affiliation{Department of Physics and Astronomy, Uppsala University, 752 36 Uppsala, Sweden}
\author {Sugata Ray}
\affiliation{Centre for Advanced Materials, Indian Association for the Cultivation of Science, Jadavpur, Kolkata 700032, India }
\affiliation {Department of Materials Science, Indian Association for the Cultivation of Science, Jadavpur, Kolkata 700032, India}

\date{\today}

\begin{abstract}
The magnetic ground state in highly ordered double perovskites LaSr$_{1-x}$Ca$_x$NiReO$_6$ ($x$ = 0.0, 0.5, 1.0) were studied in view of the Goodenough-Kanamori rules of superexchange interactions. In LaSrNiReO$_6$, Ni and Re sublattices are found to exhibit curious magnetic states, but do not show any long range magnetic ordering. The magnetic transition at $\sim$ 255 K is identified with the Re sublattic magnetic ordering. The sublattice interactions are tuned by modifying the Ni-O-Re bond angles via changing the lattice structure through Ca doping. Upon Ca doping, the Ni and Re sublattices start to display a ferrimagnetically ordered state at low temperature. The neutron powder diffraction reveals a canted alignment between the Ni and the Re sublattices, while the individual sublattice is ferromagnetic. The transition temperature of the ferrimagnetic phase increases monotonically with increasing Ca concentration.
\end{abstract}

\maketitle
\section{Introduction}
Double perovskites (DP; $A$$_2$$B$$B$$^{'}$O$_6$)~\cite{1, DP1, DP2} belong to a class of materials which exhibits many interesting properties and rich physics. Understandably, the choice of the transition metal ions at the $B$ and $B$$^{'}$ sites with different electron occupancies decide the material properties of the DPs. When both $B$ and $B$$^{'}$ are magnetic ions, the magnetic and electronic properties of the system is governed by $B$-O-$B$$^{'}$ interaction within a rock salt type structural defination as shown in Fig. 1 (a). For example, the high temperature ferromagnetic (FM) order ($T$$_C$ $>$ 400 K) of the DP compounds, Sr$_2$FeMoO$_6$ and Sr$_2$FeReO$_6$ is explained by a generalized double exchange mechanism through electronic band filling of the (Mo/Re) $t$$_{2g}$$\downarrow$-O-Fe $t$$_{2g}$$\downarrow$ conduction band~\cite{sfmo5}.
However, if the $B$-site ion is non-magnetic, the magnetic ground state would be defined by the edge-shared network of tetrahedra comprising the $B$$^{'}$ magnetic ions (Fig. 1~(b)). Such systems often exhibit geometric frustration in presence of antiferromagnetic nearest-neighbor correlations. Recently, detailed theoretical investigations have been carried out on similar DPs with the magnetic $B$$^{'}$ ions, having $n$$d$$^1$ and $n$$d$$^2$ ($n$ = 4, 5) electronic configurations and significant spin orbit coupling (SOC)~\cite{d1, d2}. Here, the nearest neighbor distance between the tetrahedrally arranged 4$d$/5$d$ magnetic ions naturally becomes much larger compared to the cases when both $B$ and $B$$^{'}$ sites are filled up with the magnetic ions. This reduces the inter-atomic exchange between the magnetic ions which helps to protect the SOC driven states. This situation opens up many options, and consequently many double perovskites with $d$$^1$ (e.g., Ba$_2$YMoO$_6$, Sr$_2$CaReO$_6$, Sr$_2$MgReO$_6$, Ba$_2$NaOsO$_6$ etc.) as well as $d$$^2$ electronic configurations (e.g., Ba$_2$CaOsO$_6$, Ba$_2$YReO$_6$, La$_2$LiReO$_6$ etc.) have been studied, where numerous unusual magnetic ground states are revealed~\cite{21, 8, 9, 10re, 11re, 12os, 13os, 14os, os-re}.

Another interesting possibility appears in DPs, when both $B$ and $B$$^{'}$ ions are magnetic, but the valence electrons of $B$ ion lack the orbital symmetry of the same of $B$$^{'}$ ion for effective $B$-O-$B$$^{'}$ superexchange interaction. Such a situation will give rise to two noninteracting or weakly interacting magnetic sublattices. It will be interesting to gradually manipulate the extent of $B$-O-$B$$^{'}$ interaction by carefully changing the $B$-O-$B$$^{'}$ angle, and consequently follow the evolution of two sublattices getting engaged in single magnetic lattice (Fig. 1 (b) $\longrightarrow$ Fig. 1 (a)), following the famous Goodenough-Kanamori rule. Accordingly, we have designed a series of DPs, LaSr$_{1-x}$Ca$_x$NiReO$_6$, having a combination of 3$d$ and 5$d$ transition metals Ni$^{2+}$ (3$d$$^8$, $t$$_{2g}$$^6$$e$$_g$$^2$) and Re$^{5+}$ (5$d$$^2$, $t$$_{2g}$$^2$) at the $B$ and $B$$^{'}$ sites, respectively. Due to large crystal field splitting, the empty $e$$_g$ orbitals of Re have much higher energy than the $t$$_{2g}$ manifold and do not contribute in any interatomic exchange interaction. The filled $t$$_{2g}$ orbital of Ni as well has less involvement in the exchange interaction. The only possible exchange interaction that can be active between the $e$$_g$ of Ni$^{2+}$ and the $t$$_{2g}$ of Re$^{5+}$, will be very weak if the $B$-O-$B$$^{'}$ angle is strictly 180$^{\circ}$, as the overlap integral between the $e$$_g$ and the $t$$_{2g}$ becomes zero. However, finite overlap between these orbitals can be introduced by tuning the bond angles and lattice parameters as a consequence of the doping of Sr$^{2+}$ ions by smaller sized Ca$^{2+}$ ions.

\begin{figure}[t]
\includegraphics[width=0.5\columnwidth]{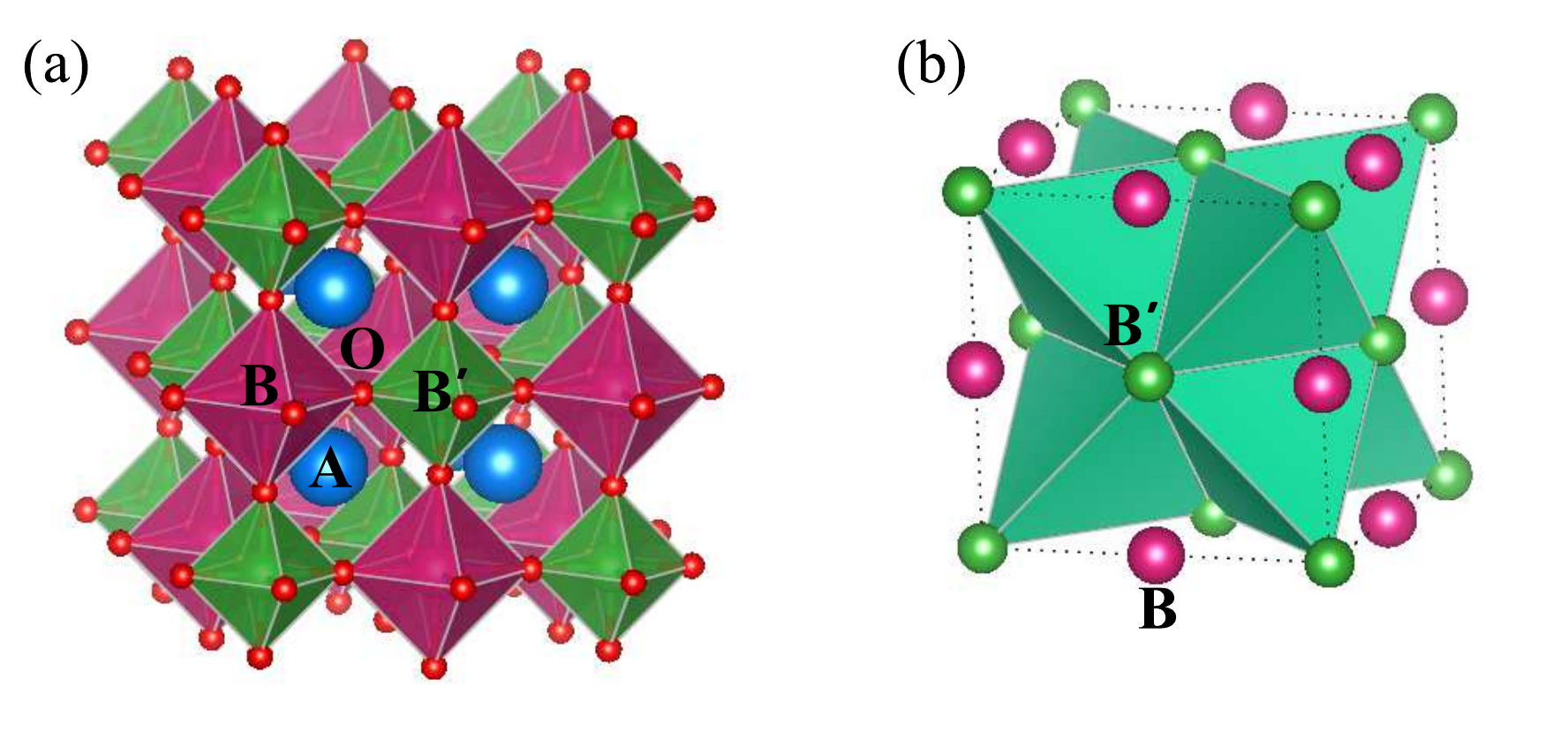}
\caption{\label{fig:1} (a) The crystal structure of the $B$-site ordered double perovskite, $A$$_2$$B$$B$$^{'}$O$_6$. (b) The geometrically frustrated face-centered cubic lattice of edge-shared tetrahedra formed by the $B$$^{'}$ sites.}
\label{Fig1}
\end{figure}

In our design criteria of the ordered LaSr$_{1-x}$Ca$_x$NiReO$_6$ series, $B$, $B$$^{'}$ cations are also selected in such a way that there is sufficiently large difference in their charge and ionic radii, in order to achieve complete $B$, $B$$^{'}$ rock-salt ordering. Here the ionic radii of Ni$^{2+}$ (= 0.69 {\AA}) and Re$^{5+}$ (= 0.58 {\AA})~\cite{shanon_5} do fulfill the above criteria. Detailed magnetic measurements on LaSr$_{1-x}$Ca$_x$NiReO$_6$ indicate curious evolution of magnetic states as a function of doping. For LaSrNiReO$_6$, the system undergoes multiple magnetic transitions, indicated by a divergence between ZFC and FC at $\sim$ 255~K, typical of Re$^{5+}$ $t$$_{2g}$$^2$ ions confined in a fcc sublattice, and by a down turn in magnetization at $\sim$ 27 K, observed in both the ZFC and FC curves. Transport measurements confirm purely insulating behavior of the samples. However with Ca doping, the structure undergoes into larger monoclinic distortion, which results in a larger deviation of the Ni-O-Re ($\angle$ NOR) bond angles from 180$^{\circ}$. This deviation enables substantial superexchange interaction between the Ni $e$$_g$ and Re $t$$_{2g}$ orbitals, resulting to an overall canted antiferromagnetic order between two individual ferromagnetic sublattices.

\begin{figure}
\begin{center}
\vspace{-0.2 in}
\resizebox{8cm}{!}
{\includegraphics[trim=50 40 180 20, clip]{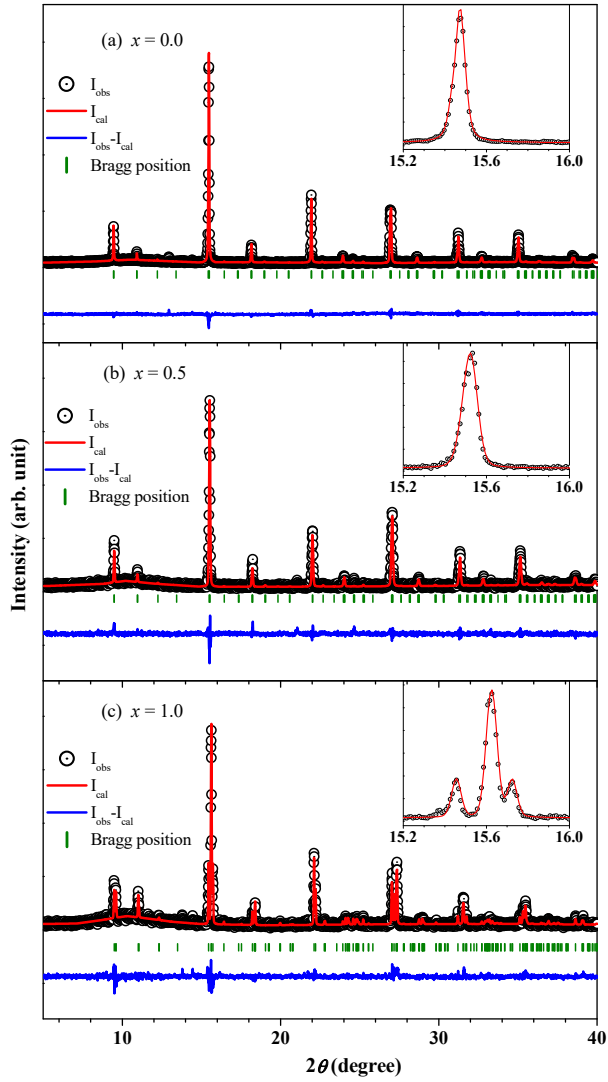}}\\
 \caption{Rietveld refinement of the X-ray diffraction pattern collected at room temperature. The observed (black circle), calculated (red line) and the difference (blue dashed line) diffraction data for (a) $x$ = 0.0, (b) $x$ = 0.5 and (c) $x$ = 1.0 compositions are displaced. Insets show the magnified view of the peaks around 2$\theta  $ $\sim$ 32$^{\circ}$. }
\vspace{-0.15 in}
 \end{center}
 \end{figure}
 
\begin{table}[h]
\begin{center}
\caption {Lattice parameters, goodness factors and the atomic parameters for $x$ = 0.0, 0.5 and 1.0 compositions obtained from Rietveld refinements.}
\vspace{0.1in}
\resizebox{8cm}{!}{%
\begin{tabular}{|c|c|c|c|c|}
\hline
\multicolumn{5}{|c|} {$x$ = 0.0; Space group = $P$2$_1$/$n$; $a$ = 5.593 \AA, $b$ = 5.571 \AA, $c$ = 7.887, \AA}\\
\multicolumn{5}{|c|} {$\beta$ = 89.996$, V$=245.72 \AA$^{3}$, R$_{wp}$(\%)=27.3, R$_{exp}$(\%)=20.49, $\chi$$^{2}$=1.78}\\
\hline
atom   & $x$ & $y$ & $z$ & Occupancy \\
\hline
Sr1/La1 & 0.499  & 0.518  & 0.247 & 0.5/0.5 \\
Ni1/Re1 &  0.500 & 0.000 & 0.500 & 0.465/0.035 \\
Re2/Ni2 &  0.500 & 0.000 & 0.000 & 0.469/0.031 \\
O1 &  0.241 & 0.237 & 0.019 & 1.0 \\
O2 & 0.285 & 0.710 & 0.978 & 1.0 \\
O3 &  0.558 & -0.008 & 0.251 & 1.0 \\
\hline
\multicolumn{5}{|c|} {$x$ = 0.5; Space group = $P$2$_1$/$n$; $a$ = 5.556 \AA, $b$ = 5.574 \AA, $c$ = 7.858 \AA, }\\
\multicolumn{5}{|c|} {$\beta$ = 89.982, $V$=243.35 \AA$^{3}$, R$_{wp}$(\%)=39.5, R$_{exp}$(\%)=35.48, $\chi$$^{2}$=1.24}\\
\hline
atom   & $x$ & $y$ & $z$ & Occupancy \\
\hline
Sr1/La1/Ca1 & 0.499  & 0.519  & 0.244 & 0.5/0.25/0.25 \\
Ni1/Re1 &  0.500 & 0.000 & 0.500 & 0.471/0.029 \\
Re2/Ni2 &  0.500 & 0.000 & 0.000 & 0.475/0.025 \\
O1 &  0.215 & 0.214 & -0.005 & 1.0 \\
O2 & 0.252 & 0.719 & 0.953 & 1.0 \\
O3 &  0.593 & -0.011 & 0.250 & 1.0 \\
\hline
\multicolumn{5}{|c|} {$x$ = 1.0; Space group = $P$2$_1$/$n$; $a$ = 5.490 \AA, $b$ = 5.585 \AA, $c$ = 7.796 \AA,}\\
 \multicolumn{5}{|c|} {$\beta$ = 90.118, $V$=239.04 \AA$^{3}$, R$_{wp}$(\%)=41.7, R$_{exp}$(\%)=33.45, $\chi$$^{2}$=1.55}\\
\hline
atom   & $x$ & $y$ & $z$ & Occupancy \\
\hline
Ca1/La1    & 0.486    & 0.536  & 0.250 & 0.5/0.5 \\
Ni1/Re1 &  0.500 & 0.000 & 0.500 & 0.486/0.014 \\
Re2/Ni2 &  0.500 & 0.000 & 0.000 & 0.485/0.015 \\
O1 &  0.309 & 0.287 & 0.056 & 1.0 \\
O2 &  0.196 & 0.787 & 0.0536 & 1.0 \\
O3 &  0.590 & -0.004 & 0.251 & 1.0 \\
\hline
 \end{tabular}
}
 \end{center}
 \end{table}

\section{Results and Discussions}
All the samples appeared to be single phased as no impurity peak is detected in the whole 2${\theta}$ range in the powder XRD data, collected at room temperature for LaSr$_{1-x}$Ca$_x$NiReO$_6$ (LSCNRO) series (see Fig. 2). The observed (circle), calculated (line through the data) and the difference (blue dashed line) diffraction data are shown in Fig. 2 for $x$ = 0.0, 0.5 and 1.0 compositions respectively, all of which could be fitted with the $P$2$_1$/$n$ space group. The system shows increase in monoclinic distortion (see the insets to Fig. 2 (a), Fig. 2 (b) and Fig. 2 (c)) with increasing Ca concentration. The observation of the monoclinic distortion in this series of compounds is in agreement with the calculated tolerance factors, which goes from $f$ = 0.98 ($x$ = 0.0) to $f$ = 0.96 ($x$ = 1.0).  An overall shift of the Bragg peaks towards higher 2${\theta}$ indicates a decrease of the unit cell parameters with increasing $x$. The refined parameters and the goodness factors are listed in Table 1. Presence of the intense peak at around 10$^\circ$ (1 1 0) indicates high degree of Ni/Re ordering within the double perovskite structure of LaSr$_{1-x}$Ca$_x$NiReO$_6$ for $x$ = 0.0, 0.5 and 1.0 compositions. The ordering between Ni and Re is quantified to be 93\%, 95\% and 97\% for $x$ = 0.0, $x$ = 0.5 and $x$ = 1.0 from the refined occupation numbers of Ni and Re at (0.5, 0.0, 0.5) and (0.5, 0.0, 0.0) crystallographic sites, respectively. The stoichiometry of Ni and Re in the compounds are probed and confirmed through ICP-OES experiments.

\begin{figure}[t]
\includegraphics[width=0.5\columnwidth]{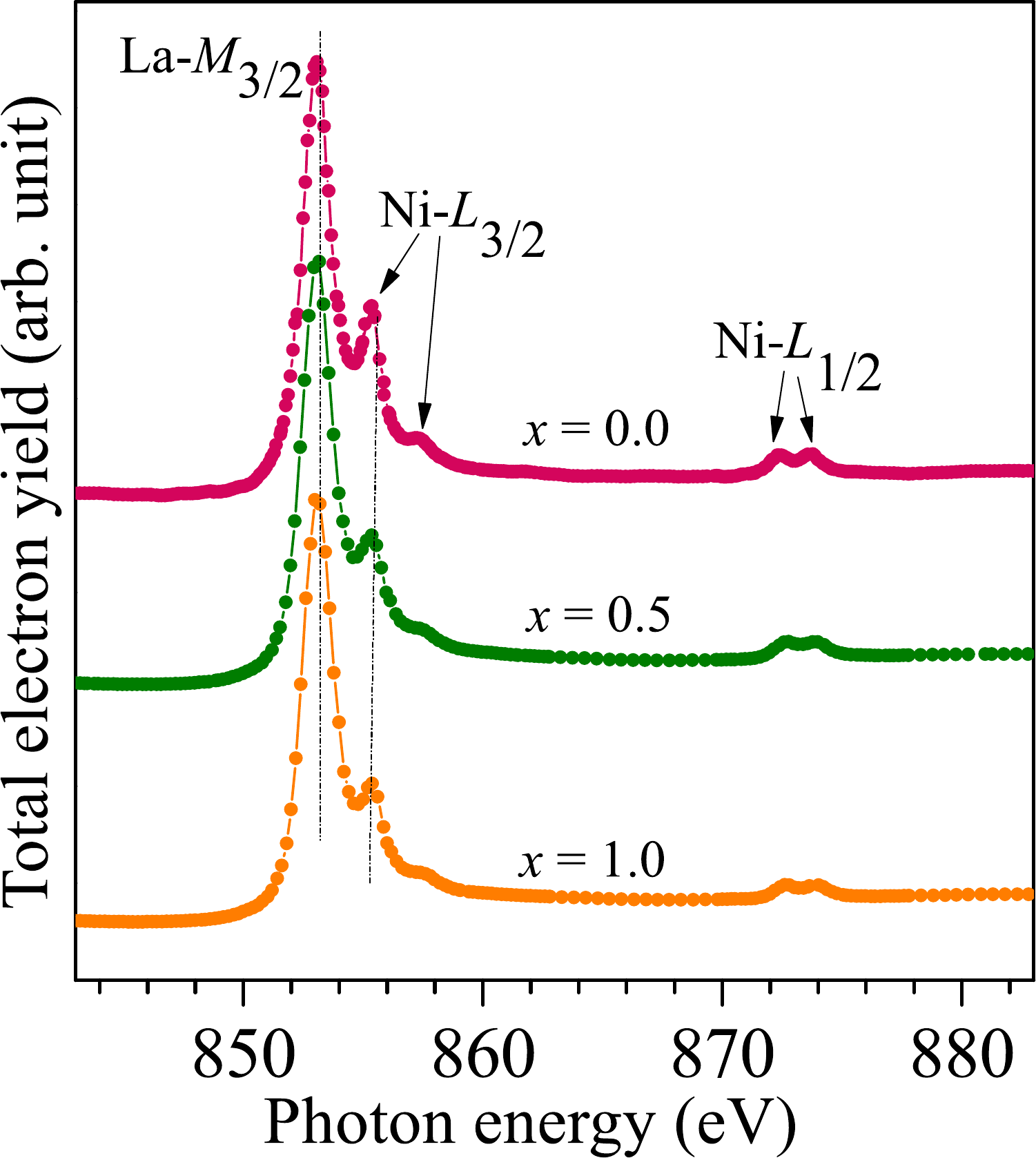}
\caption{\label{fig:3} Ni $L$-edge X-ray absorption spectra measured for $x$ = 0.0, 0.5 and 1.0 compositions of LaSr$_{1-x}$Ca$_x$NiReO$_6$ series. Data have been vertically shifted for clarity.}
\label{Fig3}
\end{figure}

\begin{figure}[t]
\includegraphics[width=0.5\columnwidth]{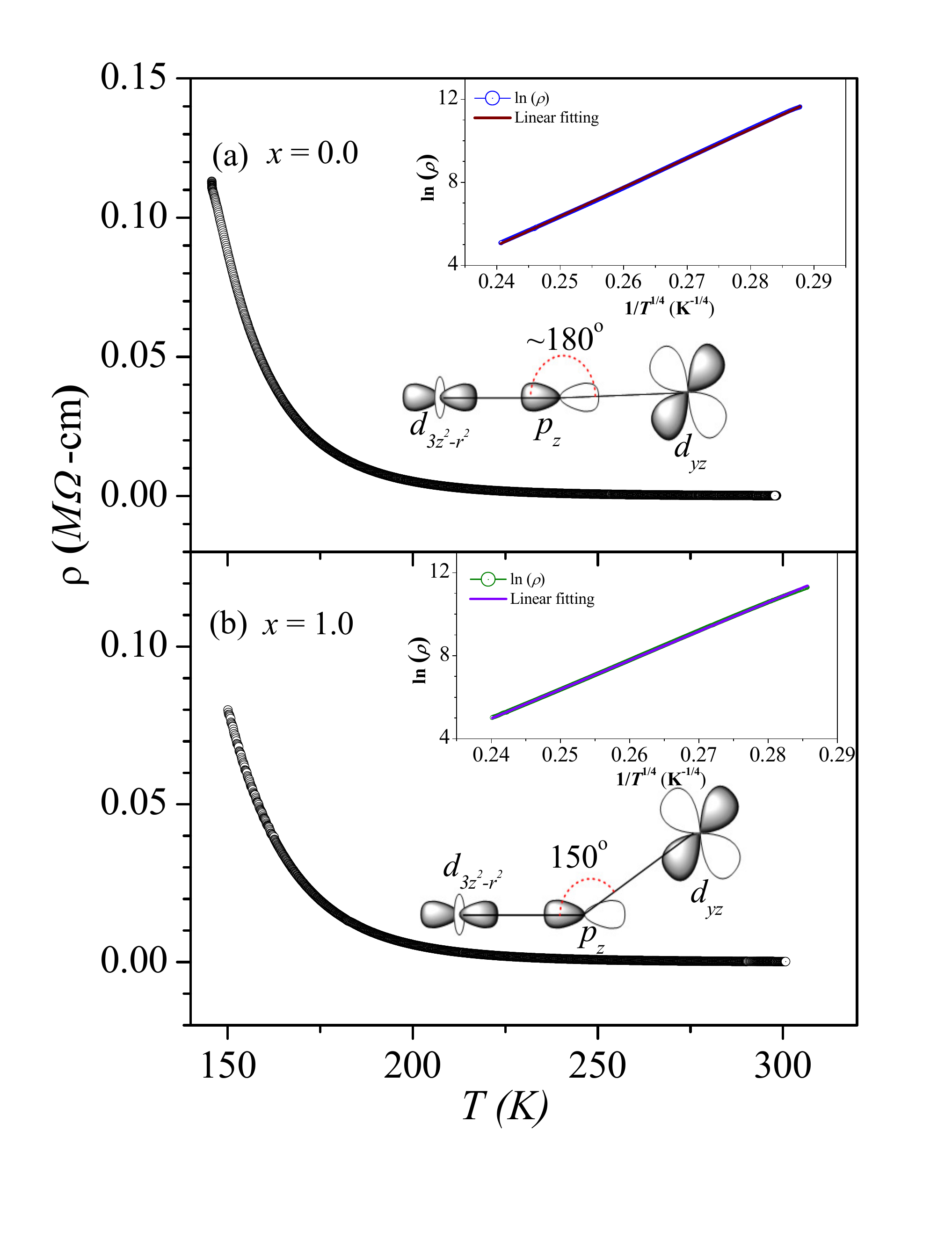}
\caption{\label{fig:4} Resistivity vs temperature plots for (a) $x$ = 0.0 and (b) $x$ = 1.0 compositions. Insets showing the insulating electrical resistivity of both the compositions could be modelled by variable range hopping mechanism in 3-dimension.}
\label{Fig4}
\end{figure}

Next, we have performed X-ray absorption spectroscopy (XAS) at the $L$-edge of Ni to verify the charge state. The XAS  spectra collected for $x$ = 0.0, 0.5, 1.0 compositions are shown in Fig. 3. Note that the La $M$$_{3/2}$ absorption edge is superimposed with the Ni $L$$_{3/2}$-edge. However, the spectral features of both the Ni $L$$_{3/2}$ and $L$$_{1/2}$ are quite different for Ni$^{2+}$ and Ni$^{3+}$~\cite{Ni_XAS}. Both $L$$_{3/2}$ and $L$$_{1/2}$ consist of two peaks indicated by the arrows in the figure. In the 2+ charge state of Ni as in NiO, the lower energy peak of $L$$_{3/2}$ is much more intense relative to the higher energy one, while both the peaks have similar intensities in case of NdNiO$_3$, PrNiO$_3$ and LaNiO$_3$, containing Ni$^{3+}$ ions~\cite{Ni_XAS}. Also, the spectral feature of the $L$$_{1/2}$ edge of all the three compositions resemble to that of NiO. Hence it can be inferred that Ni is in 2+ charge state for all the LSCNRO compositions. In order to maintain charge neutrality, the charge state of Re should then be 5+ as Sr$^{2+}$, La$^{3+}$ and O$^{2-}$ are expected to be very stable at their respective ionic states.

\begin{figure}[t]
\includegraphics[width=0.5\columnwidth]{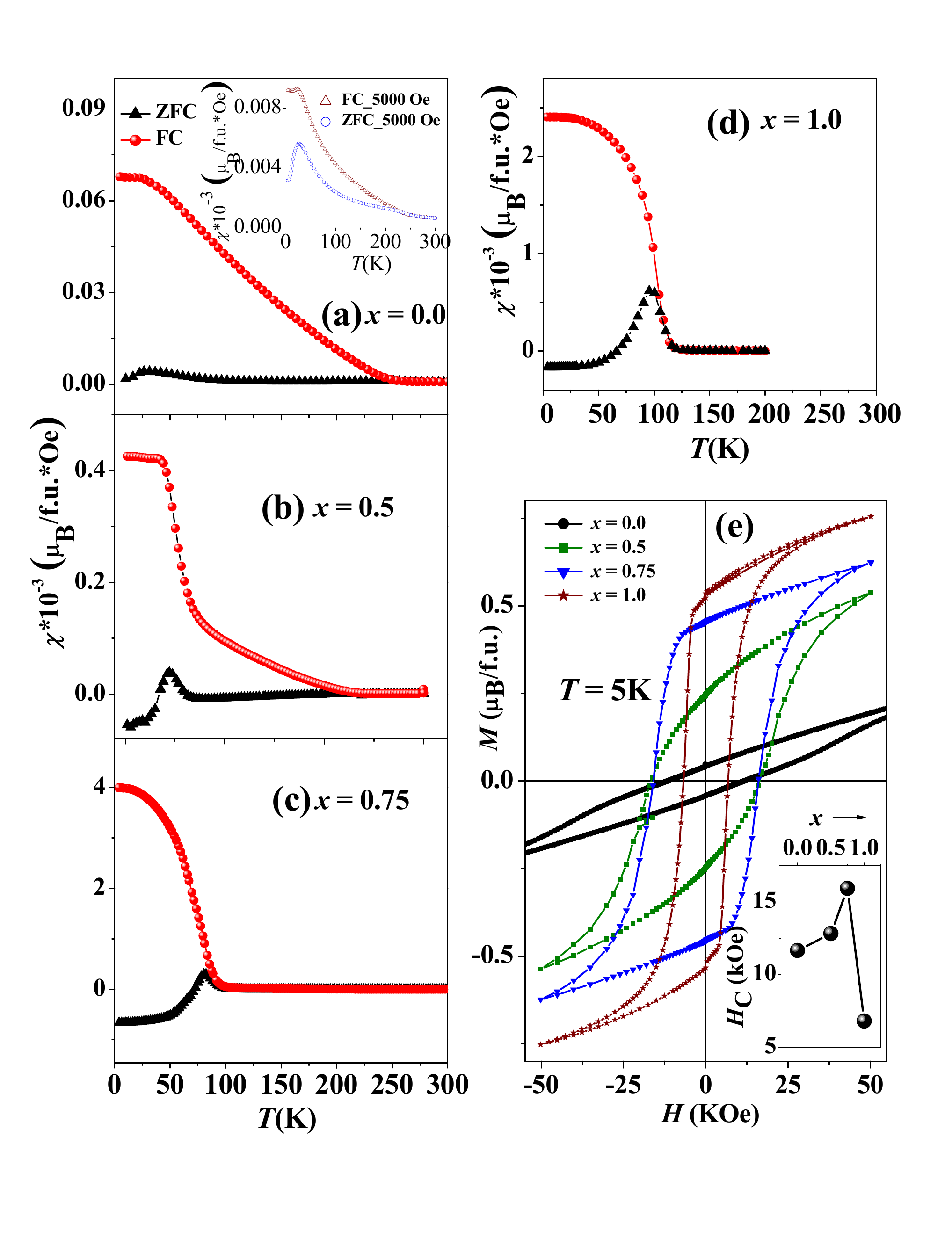}
\caption{Magnetic measurements. (a)-(d) ZFC FC, ${M(T)}$ data of LaSr$_{1-x}$Ca$_x$NiReO$_6$ with $x$ = 0.0, 0.5, 0.75 and 1.0 measured with $H$ = 200 Oe. Inset of (a) shows the temperature dependence of magnetic susceptibility of $x$ = 0.0 composition in an applied field of 5000 Oe. (e) ${M(H)}$ curves for $x$ = 0.0, 0.5, 0.75 and 1.0 compositions. Inset shows variation of $H$$_C$ with doping.}
\label{Fig5}
\end{figure}

The electrical resistivity ($\rho$) of the two end member compounds as a function of temperature are shown in Fig. 4. The resistivity data were fitted using an activated transport model as well as using variable range
hopping model. Both the data were nonlinear on a $T$$^{-1}$ scale and was found to be linear on a $T$$^{-1/4}$ scale (see inset to Figs. 4 (a) and (b)), in accordance with a three-dimensional variable range hopping transport model \cite{vth}.
The insulating nature of the compounds can be explained by the fact that Ni$^{2+}$; 3$d$$^8$ effectively provides electrons of $e$$_g$ symmetry at the valence band as the $t$$_{2g}$ bands are completely filled up, while Re$^{5+}$; 5$d$$^2$ has partially filled $t$$_{2g}$ bands, thus from the symmetry consideration the electron hopping is prohibited (see the inset of Fig. 4 (a)). However, a nonzero hopping probability is realized if the bond angles deviate sufficiently from 180$^{\circ}$, thereby enabling Ni $e$$_g$-Re $t$$_{2g}$ hybridization via oxygen (compare the inset to Fig. 4 (b)). Replacing Sr$^{2+}$ by the smaller Ca$^{2+}$ ion, we anticipate a larger octahedral distortion that can provide a route for hybridization between the Ni $e$$_g$ and Re $t$$_{2g}$. Indeed a larger amount of distortion is achieved as understood from the crystal structure of LaCaNiReO$_6$ and consequently a clear decrease in the resistivity is also observed, although the temperature dependence still suggests that the material is best described as an insulator/semiconductor.

Next, we have looked into the magnetic properties in details. The zero field cooled (ZFC) and field cooled (FC) data recorded with an applied field of 200 Oe for the four compositions of LSCNRO are shown in Figs. 5 (a)-(d). The ${M(T)}$ of $x$ = 0.0 sample shows a transition at around 255 K where ZFC and FC curves start to bifurcate. Another transition occurs at around 27 K, where susceptibilty seems to saturate.
For $x$ = 0.5 (Fig. 5 (b)) a ferro/ferrimagnetic (FM) like transition is observed at 45 K, along with the high temperature FC-ZFC bifurcated curves. This FM like transition gradually shifts to higher temperatures with increasing $x$ ($\sim$ 81 K for $x$ = 0.75 and $\sim$ 110 K for $x$ = 1.0;~Fig. 5) while the higher temperature FC-ZFC splitting continuously diminishes and in fact vanishes in case of $x$~=~1.0. It appears that the parent compound with strontium ($x$~=~0.0), hosting two separate magnetic instabilities, gets converted to a single magnetic lattice with an unique magnetic order when all the Sr ions are replaced with Ca.

The magnetization versus field ${(M (H))}$ data, collected at 5 K for the four compositions are plotted in Fig. 5 (e). The coercivity of all the samples are very high in general, which arises as a result of large spin-orbit coupling driven anisotropy, commonly observed for Re based DPs~\cite{sfmo5,Re}. However, the overall nature of the $M$($H$) curves varies drastically with $x$. The remnant magnetization increases continuously with clear signatures of magnetic saturation as a function of Ca-doping. This observation clearly suggests significant changes in magnetic interactions with increasing monoclinic distortions and decreasing $B$-O-$B$$^{'}$ angle.

\begin{figure}
\begin{center}
\vspace{-0.17 in}
\resizebox{8cm}{!}
{\includegraphics[trim=90 200 60 60, clip]{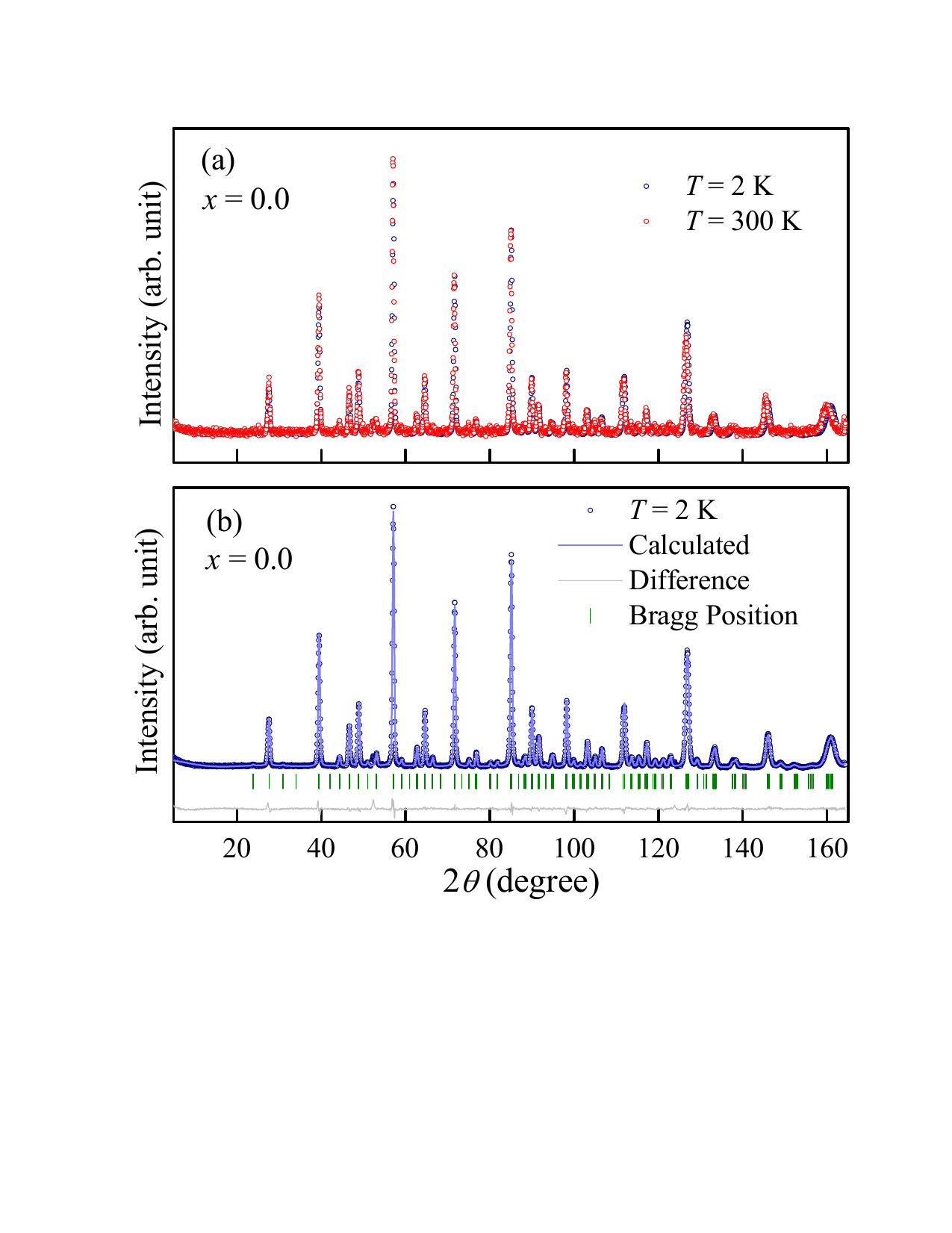}}\\
\caption{(a) Neutron diffraction pattern for $x$ = 0.0 recorded at 2 K and 300 K. No magnetic Bragg peaks are observed. (b) Rietveld refinement of NPD data at 2 K. The solid line through the experimental points is the Rietveld refinement profile calculated for the space group $P$2$_1$/$n$ structural model. The vertical bars indicate the Bragg positions. The lowermost curve represents the difference between the experimental data and the calculated results.}
\label{Fig6}
\vspace{-0.25 in}
\end{center}
\end{figure}


\begin{figure}
\begin{center}
\vspace{-0.17 in}
\resizebox{8cm}{!}
{\includegraphics[trim=30 20 0 0, clip]{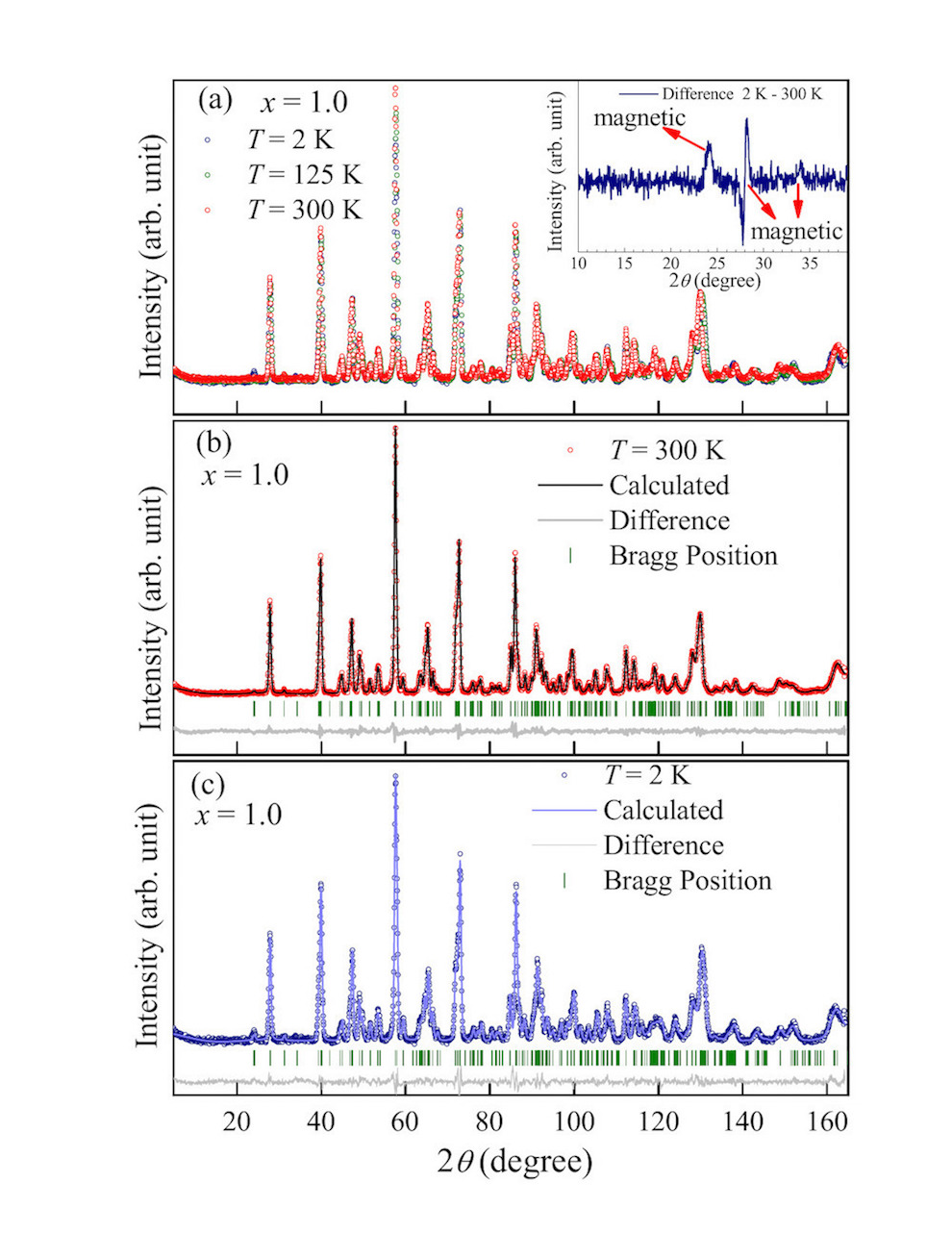}}\\
\caption{ (a) Neutron diffraction pattern for $x$ = 1.0 recorded at 2 K, 125 K and 300 K. Inset shows the magnetic Bragg peaks that are observed below the transition temperature. Rietveld refinement of NPD data for $x$ = 1.0 at (b) $T$ = 300 K and (c) $T$ = 2 K. The solid line through the experimental points is the Rietveld refinement profile calculated for the space group $P$2$_1$/$n$ structural model. The vertical bars indicate the Bragg positions. The lowermost curve represents the difference between the experimental data and calculated results.}
\label{Fig7}
\vspace{-0.25 in}
\end{center}
\end{figure}


In order to have more insight about the observed magnetic transitions, neutron powder diffraction (NPD) were carried out for the end compositions above and below the magnetic transition temperatures. Fig. 6 (a) displays the NPD patterns for $x$ = 0.0 measured at 2 K and 300 K. A comparison of the data below and above the transition temperature (see Fig. 6 (a)) reveals no magnetic Bragg peak. This suggests that the observed magnetic transitions are of short range type.
A structural Rietveld refinement of the neutron diffraction pattern at $T$ = 300 K shown in Fig. 6 (b) reveals a single-phase nature of the sample with crystallographic parameters very close to what have been extracted from the x-ray diffraction pattern. At 300 K the lattice parameters are $a$ = 5.599 \AA, $b$ = 5.575 \AA~and $c$ = 7.893 \AA.

The neutron diffraction patterns of $x$ = 1.0 recorded at 2 K, 125 K and 300 K are shown in Fig. 7 (a). Contrary to $x$ = 0.0, several magnetic Bragg peaks are clearly observed and easily identified by the difference plot shown in the inset of Fig. 7 (a). There are three clear peaks at low $q$, which are indexed with propagation vector $k$ = [0,0,0]. These peaks are not forbidden to have structural origin for the paramagnetic space group $P2_1/n$, but it will result changes in many other peaks at higher angles. Experimental intensity of the first magnetic peak (101) at 2$\theta$ $\sim$ 24$^\circ$  is not observed above $T$$_N$ at 125 K, but it is not zero by symmetry.
We have tried to completely release the structure parameters in the fit of low temperature pattern but the peak (101) is not fitted, implying that it must be of magnetic origin. We have very wide $q$-range and apparently the structural origin of (101) peak is not supported by the whole diffraction patters, there are many other Bragg peaks that would be affected by the structural change. In addition, we have performed low temperature XRD (not shown in this manuscript) and no structural change was found. A Shubnikov space solution is found to correspond to a irreducible representation mGM2+mk7t3 P1 (a) 14.79 P21'/n' with the lattice parameters  $a$ = 5.482 \AA, $b$ = 5.593 \AA~ and $c$ = 7.782 \AA~and a magnetization per unit cell of 1.067 $\mu{_B}$. The fitted parameters for the magnetization vector for both Ni and Re site are given in the Table 2. The magnetic moments at Ni and Re site are plotted in Fig. 8. Both Ni and Re moments are ferromagnetically aligned within individual sublattice, while the refined structure suggests a canting alignment between the two sublattice.

\begin{table}{}
\begin{center}
\caption {NPD results}
\vspace{0.45in}
\resizebox{12cm}{!}
{
\begin{tabular}{cccccc}
\hline
\multicolumn{6}{c}{$x$ = 1.0 (LaCaNiReO$_6$)} \\
\hline
Atom type & $M_x$ ($sM_x$) & $M_y$ ($sM_y$) & $M_z$ ($sM_z$) & $M$ ($sM$) & $M_{phase}$ ($sM_{phase}$) \\
\hline
\hline
 Ni & -1.268 (622) & -0.290 (285)  & 1.038 (540)  &  1.6664 (1830)  &  0.0000 (0)  \\
 Re & -0.670 (612) &  0.037 (281) & -0.149 (531) & 0.6865 (7072) & 0.0000 (0) \\
\hline
\end{tabular}
}
\end{center}
\end{table}

\begin{figure}[t]
\includegraphics[width=0.4\columnwidth]{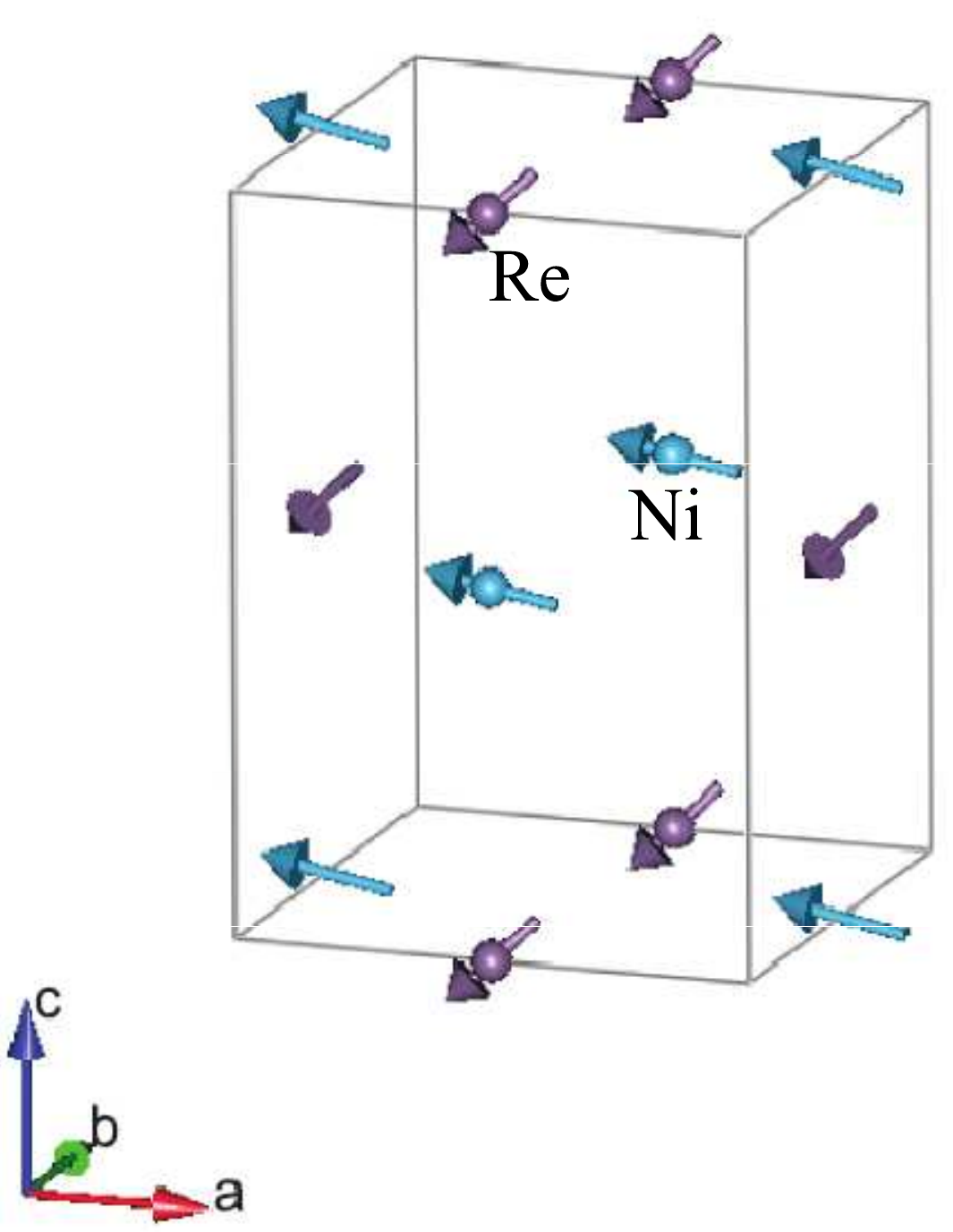}
\caption{Magnetic structure for $x$ = 1.0 at $T$ = 2 K. Arrows indicate the direction of the Ni and Re moments. The La/Ca, and O atoms are not shown for clarity. The average value of the ordered magnetic moment is estimated as 1.067 $\mu{_B}$ per unit cell.}
\label{Fig8}
\end{figure}

In a highly ordered $B$-site double perovskite with magnetic ions at $B$ and $B$$^{'}$ sites the long range magnetic order is always determined by the type of active exchange interaction that mediates through the $B$-O-$B$$^{'}$ connectivity.  For localized electrons (i.e. the $d$ electrons), the magnetic interactions between two such magnetic $B$ and $B$$^{'}$ cations is often described by the Goodenough-Kanamori rules of superexchange interaction \cite{good,kanamori}. According to this rule, when the orbitals of two magnetic ions have a significant overlap integral, the superexchange is antiferromagnetic ($\angle$ $e$$_g$($B$)-O-$e$$_g$($B$$^{'}$) = 180$^{\circ}$,  $\angle$ $t$$_{2g}$($B$)-O-$t$$_{2g}$($B$$^{'}$) = 180$^{\circ}$, $\angle$ $t$$_{2g}$($B$)-O-$e$$_g$($B$$^{'}$) = 90$^{\circ}$). However, when the orbitals are arranged in such a way that they are expected to be in contact but to have no overlap integral - most notably $t$$_{2g}$ and $e$$_g$ in 180$^{\circ}$ position, where the overlap is zero by symmetry, the rules predict ferromagnetic interaction, which is usually very weak in strength. In case of highly ordered LaSr$_{1-x}$Ca$_x$NiReO$_6$ compounds, Ni 3$d$ and Re 5$d$ orbitals are connected by oxygen 2$p$ orbitals. From refinement of XRD, the average Ni-O-Re bond angle ($\angle$ NOR) comes about 170$^{\circ}$  for $x$ = 0.0 sample and $\sim$150$^{\circ}$ for $x$ = 1.0 sample. Therefore in case of $x$ = 0.0 sample, when $\angle$ NOR is close to 180$^{\circ}$, the only superexchange interaction one can expect between half filled $e$$_g$ orbitals of Ni$^{2+}$ and partially filled $t$$_{2g}$ orbitals of Re$^{5+}$ is a weak ferromagnetic interaction. Of course, there could be magnetic signals appearing from independent Ni and Re sublattices too. However, as the $\angle$ NOR start deviating noticeably from 180$^{\circ}$ (e.g. $x$ = 1.0), the nonzero overlap between the $e$$_g$ and $t$$_{2g}$ orbitals starts to favor antiferromagnetic (AFM) interaction between the partially filled Ni $e$$_g$ and Re $t$$_{2g}$ orbitals following the Goodenough-Kanamori rule. We conclude that the low temperature magnetic feature observed in $x$~=~0.0 compound could very well be a reminiscence of the weak ferromagnetism predicted by Goodenough-Kanamori rule, but could also be an independent Ni sublattice feature as is seen in other Ni analog samples, such as, Sr$_2$NiWO$_6$ and Sr$_2$NiTeO$_6$ with nonmagnetic W$^{6+}$ (${[Xe]}$4$f$$^{14}$) and Te$^{6+}$ (${[Kr]}$4$d$$^{10}$) at the $B$$^{'}$ site respectively, where the antiferromagnetic transition occurs at 35 K and 54 K respectively~\cite{ni}. However, the observed bifurcation between ZFC and FC curves at higher temperatures in $x$~=~0.0 compound is clearly very similar to what is commonly observed when Re-ion sits in the geometrically frustrated fcc lattice (the $B$$^{'}$-site ) within the double perovskite structure, but having nonmagnetic $B$-site ions, e.g., Sr$_2$CaReO$_6$~\cite{10re}, Sr$_2$InReO$_6$~\cite{siro}.

For $x$ = 1.0 sample, a sizeable deviation of $\angle$NOR from 180$^{\circ}$ enables the antiferromagnetic interaction between Ni$^{2+}$ and Re$^{5+}$ ions. In an ordered structure, this will result the moment of the individual sublattice to order along the same direction. 
However, the alignment between the two sublattice will depend on the different competitive interaction strengths. The spin-orbit interaction will result different crystalline anisotropy directions for Re and Ni spins due to the $t$$_{2g}$ and $e$$_g$ type orbitals, respectively. Therefore, the final magnetic structure is resulted in a canted arrangement between the two subllatice. In LSCNRO, both Ni$^{+2}$ and Re$^{+5}$ ions have two unpaired electrons. However, the strong spin-orbit coupling in 5$d$ orbitals (relative to 3$d$ orbitals) usually results in a reduced total moment in Re ions compared to its spin only value~\cite{Ref,Ref2,Ref3}, which is clearly observed from the NPD analysis. Also, the net magnetic moment obtained from NPD is in very good agreement with the observed moment in MH data at the highest applied field.

\section{Conclusion}
Double perovskite series, LaSr$_{1-x}$Ca$_x$NiReO$_6$ ($x$ = 0.0, 0.5, 1.0) is realized with partially filled orbitals of $e$$_g$ and $t$$_{2g}$ symmetries (local) at highly ordered $B$ and $B$$^{'}$-sites respectively. All the compositions are formed in a monoclinic structure. In LaSrNiReO$_6$ ($x$ = 0.0), an unusual divergence between the ZFC and FC curves is identified with the magnetic state that arise for Re $t$$_{2g}$$^{2}$ ions sitting in a geometrically frustrated fcc sublattice in DP host. At low temperature ($\sim$ 27 K), the system undergoes into another magnetic transition, where weak ferromagnetism predicted by Goodenough-Kanamori rule could be identified. As lattice parameter decreases with Ca doping, the reduced Ni-O-Re bond angle introduces nonzero overlap integral between Ni $e$$_g$ and Re $t$$_{2g}$  orbitals, which favors a highly canted AFM alignment between Ni and Re sublattices. The neutron powder diffraction measurement conducted at room temperature and low temperature (2 K) revealed that the $x$ = 0.0 sample possess a disordered/short-range magnetic state at low temperature, while for Ca sample, the Ni and Re sublattice aligned in a canted antiferromagnetic state to give a long range magnetic order evidenced from the magnetic Bragg peak corresponding to the double perovskite superlattice peak.

\section{Methods}
Four samples of LaSr$_{1-x}$Ca$_x$NiReO$_6$ (LSCNRO) ($x$ = 0.0, 0.5, 0.75, 1.0) were synthesized by solid state synthesis route. Highly pure La$_2$O$_3$, SrCO$_3$, CaCO$_3$, NiO, Re$_2$O$_7$ and Re metal were used as the starting materials. The synthesis was done in two steps. In the first step, La$_2$NiO$_4$ was made by heating a thorough mixture of stoichiometric La$_2$O$_3$ and NiO at 1250$^{\circ}$~C in an inert atmosphere for 48 hours with several intermediate grinding. SrO and CaO were used after heating them at 1250$^{\circ}$~C and 1000$^{\circ}$~C for 12 hours in inert atmosphere. Next, stoichiometric amount of La$_2$NiO$_4$, SrO, CaO, NiO, Re$_2$O$_7$ and Re metal were mixed inside a glovebox and the resultant mixture was sealed inside a quartz tube, which was then annealed at 1200$^{\circ}$~C for the final product.

The phase purity of the three samples ($x$ = 0.0, 0.5, 1.0) were checked by x-ray diffraction (XRD) at MCX beamline of the Elettra Synchrotron Centre, Italy using wavelength of 0.751 {\AA}. The XRD data were analyzed via Rietveld refinement using the FullProf~\cite{fullprof_5} program. La, Ca, Sr, Ni, Re quantitative analysis were performed in Inductively Coupled Plasma Optical Emission Spectroscopy (ICP-OES)(Perkin-Elmer USA, Optima 2100 DV) instrument following standard protocol of sample analysis. $d.c.$ magnetic measurements were carried out in a Quantum Design SQUID magnetometer. Resistivity measurements were performed in a home made four probe setup. Soft x-ray absorption spectroscopy (XAS) was performed at I1011 beamlines of the Swedish synchrotron facility MAX-lab, Lund. All the XAS spectra were measured by recording the total electron yield. Neutron powder diffraction (NPD) measurements were performed using  the HRPT \cite{Fischer}  diffractometer at the Paul Scherrer Institut, SINQ (Switzerland). The neutron wavelength was set to $\lambda$ = 1.89 \AA~and about 1 g of $x$ = 0.0 and $x$ = 1.0  samples were used. Magnetic structure refinements were performed using the FULLPROF suite \cite{fullprof_5}.


\section{Acknowledgement}
S.J. and P.A. thanks Council of Scientific and Industrial Research (CSIR), India for fellowship. P.A.K. thanks the Swedish Foundation for International Cooperation in Research and Higher Education (STINT) for supporting his stay at Uppsala University. S.R. thanks Indo-Italian POC for support to carry out experiments (20145381) in Elettra , Italy. S.R. also thanks Department of Science and Technology (DST) [Project No. WTI/2K15/74] and TRC, Department of Science and Technology (DST), Government of India for support.
MM and OKF are supported by a Marie Sk{\l}odowska Curie Action, International Career Grant through the European Union and Swedish Research Council (VR), Grant No. INCA-2014-6426. EN is fully funded by the Swedish Foundation for Strategic Research (SSF) within the Swedish national graduate school in neutron scattering (SwedNess). Finally, YS is fully supported by a VR neutron project grant (BIFROST, Dnr. 2016-06955) as well as a VR starting grant (Dnr. 2017-05078).The experimental neutron diffraction work was performed at the HRPT beamline of the Laboratory for Neutron Scattering \& Imaging, Paul Scherrer Institute, CH-5232 Villigen PSI, Switzerland. We thank the beamline staff for their support.

\section{Author contribution statement}
S.J. designed the project and the method of synthesis. S.J. and P.A. have synthesized and characterized the compounds. S.J. and P.A. have performed the XAS and R vs T measurements and analysis. The magnetic measurements and analysis are performed by S.J., P.A., P.A.K. and P.S., while O.K.F., E.N., V.P., M.M. and Y.S. have performed the NPD measurements and analysis. S.J., P.A. and S.R. wrote the main body of the manuscript, while all the authors contributed in writing and reviewing the manuscript.

\section{Additional Information}
{\bf Competing interests:} The authors declare no competing financial and non-financial interests. 

\end{document}